\documentclass[3p,times,twocolumn,numbers,sort&compress]{elsarticle}
\usepackage[T1]{fontenc}
\usepackage[utf8]{inputenc}
\usepackage{amsmath,amssymb}
\usepackage{graphicx}
\usepackage{float}
\usepackage{color}
\usepackage{bm}
\usepackage{hyperref}
\usepackage{booktabs}
\usepackage{multicol}
\usepackage{lipsum}
\usepackage{array}
\usepackage{caption}
\usepackage{siunitx}
\usepackage{multirow}

\usepackage{flushend}
\usepackage{balance}

\journal{Journal of Magnetism and Magnetic Materials}

\begin{document}

\begin{frontmatter}

\title{Magnetization relaxation of interacting chains of nanomagnets}

\author[1]{D. Ledue\corref{cor1}}
\cortext[cor1]{Corresponding author}
\ead{denis.ledue@univ-rouen.fr}
\author[1]{R. Patte}
\author[2]{F. Vernay}
\author[2]{H. Kachkachi}

\address[1]{Normandie Université, UNIROUEN, INSA Rouen, CNRS, Groupe de Physique des Matériaux (UMR-6634), F-76800 Saint-Étienne du Rouvray, France}
\address[2]{Laboratoire PROMES CNRS (UPR-8521) \& Université de Perpignan Via Domitia, Rambla de la thermodynamique, Tecnosud, F-66100 Perpignan, France}

\begin{abstract}
We investigate the magnetization dynamics crossover from single-particle to collective behavior in a one-dimensional chain of dipolar-coupled nanomagnets with uniaxial anisotropy. Using both an intermediate-to-high damping (IHD) analytical approach based on Langer's theory and time-quantified Monte Carlo (TQMC) simulations, we derive and validate semi-analytical expressions for the relaxation rate and the magnetization relaxation curves. Our main results include: (i) a closed-form expression for the relaxation rate accounting for (weak) dipolar interactions, (ii) a two-exponential semi-analytical formula for the magnetization dynamics $m(t)$ of an interacting chain, and (iii) a systematic comparison with TQMC simulations, showing good agreement for a wide range of parameters. The analysis reveals a field-controlled crossover from uniform (macrospin-like) reversal to edge-nucleation propagation, driven by the spatial inhomogeneity of dipolar stabilization. The derived expressions provide a computationally efficient framework for predicting the relaxation behavior of dipolar-coupled nanomagnetic assemblies.
\end{abstract}


\end{frontmatter}

\section{Introduction}
The dynamic behavior of nanomagnets (NM) depends on a complex interplay
between various factors, including intrinsic properties such as size,
shape, underlying lattice structure, and energy parameters, as well
as collective effects governed by inter-particle interactions, particularly
dipolar coupling \cite{Denisov2004Thermal, RUSSIER201650}. These dipolar interactions (DI) 
strongly affect the relaxation time of assemblies of NM, thereby influencing their overall
dynamical response and stability.

One of the central challenges in optimizing NM assemblies 
for practical applications---such as catalysis, magnetic hyperthermia
\cite{Respaud_nanocube2_2010JMMM, Mehdaoui_prb2013}, magnetic resonance imaging \cite{Them_2017},
high-density recording \cite{Arai_2013,WinklerEtAl_jmc22},
and rare-earth-free permanent magnets \cite{MaurerEtal_apl91}---is
understanding and controlling the inter-nanoparticle interactions
that determine their collective properties \cite{Komogortsev2005Magnetization}.

Technologically, many challenges remain unresolved, notably how the
unique nanoscale magnetic properties of individual nano-objects can
be preserved or tuned when assembled into macroscopic systems. Fundamentally,
this transition involves the competition between intrinsic parameters---anisotropy,
size, and thermal stability---and collective effects, such as dipolar
fields and spatial organization. A key example concerns the dynamical
crossover from single-particle to collective behavior, which remains
an open question, especially regarding the onset of spin-glass-like
or superferromagnetic order depending on particle parameters\cite{Petracic01012002, NAKAMAE2014225, high_lisiecki_2023}. Another
long-standing issue is how DI modifies
the effective energy barriers and magnetization reversal times of
a single NM \cite{Denisov2004Thermal}.

Although chains of NM were once regarded as primarily theoretical
constructs, recent advances in synthesis and the discovery of magnetotactic
bacteria have led to the experimental realization of 1D magnetic assemblies.
These systems hold promise for biomedical and technological applications,
notably magnetic hyperthermia \cite{AlphanderyEtAl_ACS2011,Fdez-GubiedaEtAl_JAP2020}.
While their static properties are now well understood \cite{miyasaka2009slow,melo2017analysis}, the dynamical
response of such chains---particularly under competing magnetic and
dipolar fields---remains incompletely described.

A number of prior works have addressed the dynamics
of interacting NM chains using both analytical and numerical
methods. Analytical treatments often rely on mean-field or perturbative
expansions in the dipolar coupling parameter ($\xi$), leading to
closed-form approximations for the effective relaxation rate as a
function of the effective anisotropy-energy barrier ( $\sigma$) and
$\xi$ (e.g., \cite{ManishEtAl_prb19}). Complementary numerical
techniques, particularly kinetic Monte Carlo (kMC) \cite{Brinis_2014} and stochastic
Landau--Lifshitz--Gilbert (LLG) simulations, have been used to model
thermally activated magnetization dynamics and hysteresis in interacting
chains, elucidating how relaxation times and blocking temperatures
depend on $\xi$, chain length, and field orientation\cite{igllab04prb,SalvadorEtAl_physb23}.
Recent Langevin-dynamics-based approaches further resolve the full
magnetization trajectories in interacting ensembles, yet direct, quantitative comparison between analytical and numerical
frameworks remain rare.

In this context, the present work introduces a systematic
comparison between the intermediate-to-high (IHD) analytical formulation,
derived from Langer's approach, and time-quantified Monte Carlo (TQMC)
simulations applied to the magnetization dynamics of an interacting NM chain. 
Numeric and analytical approaches are indeed complementary in the sense that
the IHD formulation relies on a perturbative expansion in the DI parameter $\xi$ and
assumes that a single (quasi-uniform) mode is dominant in the reversal channel, whilst TQMC
inherently captures all possible dynamics without such restrictions. Therefore, this dual analysis enables an assessment
of the range of validity of perturbative analytical models against full stochastic dynamics 
under both magnetic and dipolar fields. By explicitly incorporating finite-temperature
effects and employing the perturbative expansion of the magnetization,
the IHD formulation extends conventional low-field approximations
and captures the modulation of dipolar coupling by thermal fluctuations.
Together, these analytical and numerical results provide a unified
framework to interpret the temperature-dependent relaxation dynamics
in dipolar-coupled nanomagnetic assemblies, bridging the gap between
single-particle and collective regimes.

Moreover, the analytical expressions derived in this work offer a
practical and computationally efficient means to explore the influence
of DI on magnetization dynamics, without resorting
to the heavy numerical cost of full-scale stochastic simulations.
While the validity of the analytical treatment is naturally bounded
by its perturbative assumptions, its transparent physical content
and predictive capability make it a valuable tool for interpreting
relaxation phenomena in interacting nanoparticle systems.

The rest of the paper is organized as follows. In Sec.~\ref{sec:Presentation-of-the} 
we present the model of a dipolar-coupled chain of NM and describe the two methods used: 
the analytical IHD approach based on Langer's theory and the TQMC method. 
In Sec.~\ref{sec:results} we report the main results: the analytical expression for the relaxation 
rate of the interacting chain, its validation by TQMC simulations, and the semi-analytical 
description of the magnetization relaxation curves. We also discuss the crossover from uniform 
to edge‑nucleated reversal revealed by correlation‑length analysis. 
Finally, Sec.~\ref{sec:conclusion} summarizes the principal findings and suggests directions for future work.

\section{Model and methods\label{sec:Presentation-of-the}}

\subsection{Model}

\begin{figure}[H]
\begin{centering}
\includegraphics[width=0.6\columnwidth]{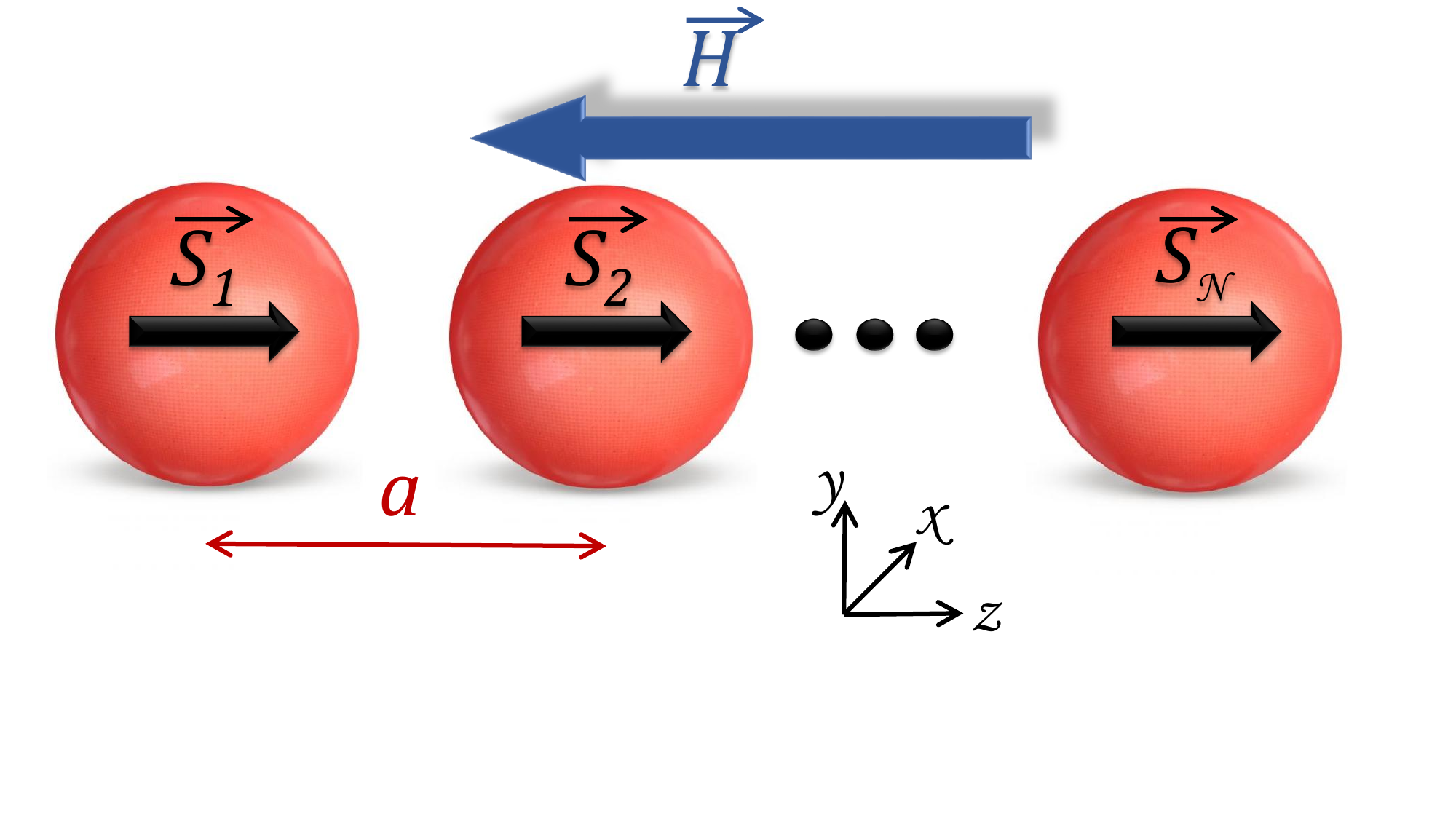}
\par\end{centering}
\caption{\label{fig:1D-chain}1D chain of $\mathcal{N}$ nanomagnets of net
magnetic moments $\bm{m}_{i}$. The nanomagnets have their uniaxial
anisotropy axes along the $z$-axis, and are subjected to a longitudinal
magnetic field (along the chain axis).}
\end{figure}

In this work, we consider the system depicted in Fig. \ref{fig:1D-chain},
which consists of a chain of $\mathcal{N}$ monodomain, monodisperse NM 
separated by an inter-particle distance
$a$. Each NM is modeled as a macrospin, carrying a net magnetic
moment $\bm{m}_{i}=m{\bf S}_{i}=n\mu_{B}{\bf S}_{i}$,
where $n$ is the number of Bohr magnetons, and ${\bf S}_{i}$ is
a unit vector. The NM possess uniaxial anisotropy with the
easy axis aligned along the chain's axis. The entire system is exposed
to an external DC magnetic field ${\bf H}$, which is oriented
along ${\bf e}_{z}$, as shown in Fig. \ref{fig:1D-chain}.

The total energy of the chain is given by the addition of the anisotropy
energy, the Zeeman energy, and the DI energy :
\begin{equation}
E=E_{{\rm A}}+E_{{\rm Z}}+E_{{\rm DI}}\label{eq:energy}
\end{equation}
where
\[
E_{{\rm A}}=-\sum_{i=1}^{\mathcal{N}}KV\left(S_{i}^{z}\right)^{2},\qquad E_{{\rm Z}}=-\sum_{i=1}^{\mathcal{N}}{\bf H}\cdot\bm{m}_{i}
\]

\begin{align*}
E_{{\rm DI}} & =-\frac{\mu_{0}}{4\pi}\frac{m^{2}}{a^{3}}\sum_{i=1}^{\mathcal{N}}\sum_{i<j}{\bf S}_{i}\mathcal{\bm{D}}_{ij}{\bf S}_{j}
\end{align*}
with
\begin{equation}
\mathcal{\bm{D}}_{ij}=\frac{3\,{\bf e}_{ij}\otimes{\bf e}_{ij}-\mathbf{I}}{r_{ij}^{3}},\quad r_{ij}=\left\Vert {\bf r}_{i}-{\bf r}_{j}\right\Vert ,{\bf e}_{ij}=\frac{{\bf r}_{ij}}{r_{ij}},\label{eq:DItensor}
\end{equation}
which means that $r_{ij}$ is unitless, the distance separating particles $i$ and $j$ being $d_{ij}=a r_{ij}$. 
The total energy can be expressed in reduced units with respect to the anisotropy field $2KV$. In the remaining of the paper, energies will be noted in simple capital letters ({\emph{e.g.}} $E$), while energies in reduced units will be in calligraphic capital letters ({\emph{e.g.}} $\mathcal{E}$). This means, for instance
\[
\mathcal{E}=E/\left(2KV\right).
\]
Accordingly, we define the dimensionless physical parameters

\begin{equation}
h\equiv\frac{H}{H_{a}};\quad\xi\equiv\frac{\mu_{0}}{4\pi}\frac{m^{2}/a^{3}}{2KV}.\label{eq:reduced_param}
\end{equation}

with
\[
H_{a}=\frac{2KV}{m}
\]

Consequently, we write

\begin{equation}
\mathcal{E}=\sum_{i}^{\mathcal{N}}\mathcal{E}_{i}=-\frac{k}{2}\sum_{i=1}^{\mathcal{N}}S_{i,z}^{2}-\sum_{i=1}^{\mathcal{N}}\bm{h}\cdot{\bf S}_{i}-\xi\sum_{i=1}^{\mathcal{N}}\sum_{i<j}{\bf S}_{i}\mathcal{\bm{D}}_{ij}{\bf S}_{j},\label{eq:total_energy}
\end{equation}
where, in these reduced units, the parameter $k$ is equal to 1, and it is introduced to keep track of the uniaxial anisotropy contribution.

The DI are directional, which makes the estimation
of the relaxation rate or other observables more complicated in the
general situation. However, in the case of a chain, only one direction
is relevant, making the calculation of the lattice sum straightforward.
We introduce the local (spherical) coordinates $\left(\theta_{i},\varphi_{i}\right)$
for each spin ${\bf S}_{i}$, with $S_{i,z}\equiv\cos\theta_{i}$.
We also define the compact notation $V_{ij}\equiv V\left(\bm{r}_{i}-\bm{r}_{j}\right)=1/\left\Vert \bm{r}_{i}-\bm{r}_{j}\right\Vert ^{3}$.
Then, owing to the symmetry with respect to rotations about the chain's
axis ($z$), the local energy becomes

\begin{equation}
\mathcal{E}_{i}=-\bm{h}\cdot\bm{S}_{i}-\frac{k}{2}S_{i,z}^{2}-\xi{\displaystyle \sum_{j\ne i}}V_{ij}\left[2S_{i,z}S_{j,z}-{\displaystyle \sum_{\alpha=x,y}}S_{i,\alpha}S_{j,\alpha}\right].\label{eq:EnergyLocal}
\end{equation}

\subsection{Relaxation rate within Langer's approach}

\begin{figure}[H]
\begin{centering}
\includegraphics[scale=0.5]{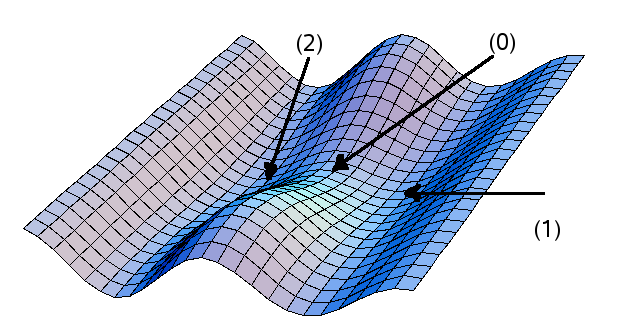}
\par\end{centering}
\caption{\label{fig:Double-well-potential}Double-well potential.}
\end{figure}

The relaxation rate (or inverse relaxation time) for a single magnetic
moment is the probability of its escape (per unit time) from an energy minimum (1)
to a minimum (2) through a saddle point (0), see Fig. \ref{fig:Double-well-potential}.

In the case of high energy barriers $\Delta E$, \emph{i.e.}, when
$\beta\Delta E\gg1$ (with $\beta=1/k_BT$) and also IHD, one may apply Langer's approach
\cite{lan68prl,lan69ap} which yields the relaxation rate in the
form \cite{kac03epl,kac04jml}
\begin{equation}
\Gamma=\frac{\left|\kappa\right|}{2\pi}\frac{\tilde{Z}_{s}}{Z_{m}},\label{eq:LangerRR}
\end{equation}
where $Z_{m}$ and $\tilde{Z}_{s}$ are respectively the partition
functions computed in the vicinity of the minimum (1) and the saddle point (0) 
and $\left|\kappa\right|$ is the attempt frequency. In practice,
for a given elementary process, \emph{i.e.} an escape from the minimum
$(\theta^{(m)},\varphi^{(m)})$ through the saddle point $(\theta^{(s)},\varphi^{(s)})$,
we have to compute the partition function
\[
Z=\int\left(\mathcal{D}\mathbf{s}\right)\,e^{-\beta E\left(\mathbf{s}\right)}
\]
at the saddle point and at the minimum. For this, one performs a quadratic
expansion of the energy at these stationary states. This is where
Langer's approach meets its limit of validity because such an expansion
is only meaningful when the stationary point is well defined.

Note that in a general situation, for instance when a magnetic field
is applied in an arbitrary direction, the loci of the stationary points
must be found numerically. Accordingly, the relaxation rate can also
be calculated numerically using the matrix-continued fraction method
as suggested by Risken \cite{risken96springer} and developed in
\cite{cofkalwal05WS}, which turns the problem of solving the Fokker-Planck
equation into an eigenvalue problem for which the smallest non-vanishing
eigenvalue $\lambda_{1}$ yields the relaxation rate $\Gamma=\lambda_{1}$.
However, this method becomes rather cumbersome beyond the macrospin
problem because one has to build a hierarchy of equations for each
energy potential, see \emph{e.g.} Ref. \cite{titovetal05prb} for
the two-spin problem. There are at least two other numerical methods
which are more versatile in dealing with multi-variate spin systems,
such as the solution of the Landau-Lifshitz equation in the stochastic
approach \cite{lybcha95prb,garpal00acp}, and the TQMC method \cite{nowaketal00prl,ChubykaloEtAl_prb03}. 

In the present work, we use Langer's approach \cite{lan68prl, lan69ap} to derive (approximate)
analytical expressions for the relaxation rate of a chain of NM
coupled by DI, under appropriate assumptions as
discussed in the sequel. We then compare the results with our TQMC simulations.

\subsection{Time-step quantified Monte Carlo method}

To numerically compute the relaxation time $\tau_{-}=1/\Gamma_{-}$, we
use the TQMC method \cite{nowaketal00prl,ChubykaloEtAl_prb03}
combined with the heat-bath algorithm, where the MC step is time quantified
according to the relation

\begin{equation}
\delta t=\frac{(1+\alpha^{2})mR^{2}}{20\alpha\gamma k_{B}T},\label{eq:Deltat_MC}
\end{equation}
where \textit{$\gamma=1.76\times10^{11}{\rm (T.s)^{-1}}$} is the
gyromagnetic ratio.\textit{ }$0<R<1$\textit{ }is the radius of a
sphere in which a random vector is chosen with a uniform probability
distribution and added to the unit macrospin $\bm{S}_{i}$ to find
the new orientation of the macrospin (of the nanoparticle). The expression
above is only valid under the condition
\begin{equation}
 R^{2}\ll\frac{3.5}{\sigma(1-h)},\label{eq:cone-radius}
\end{equation}

where
\begin{equation}
 \sigma = \frac{KV}{k_B T}.\label{eq:sigma}
\end{equation}

In the case of high damping, the 
TQMC method yields very similar results to the Langevin dynamics, \emph{i.e.}
the numerical solution of the Landau-Lifshitz equation in the Langevin
approach \cite{ChubykaloEtAl_prb03}. The
agreement between the two methods improves when the direction of the
applied field is close to the easy axis.

In the initial state, as shown in Fig. \ref{fig:1D-chain}, all magnetic moments are in the metastable state,
namely along the $+z$ axis in the case of a longitudinal field in the $-z$ direction. These minima apply in the absence of DI.
Next, each magnetic moment is set free to fluctuate in a random motion
within the potential well before it can reverse its direction after
a given time.

\begin{figure}[H]
\begin{centering}
\includegraphics[scale=0.5]{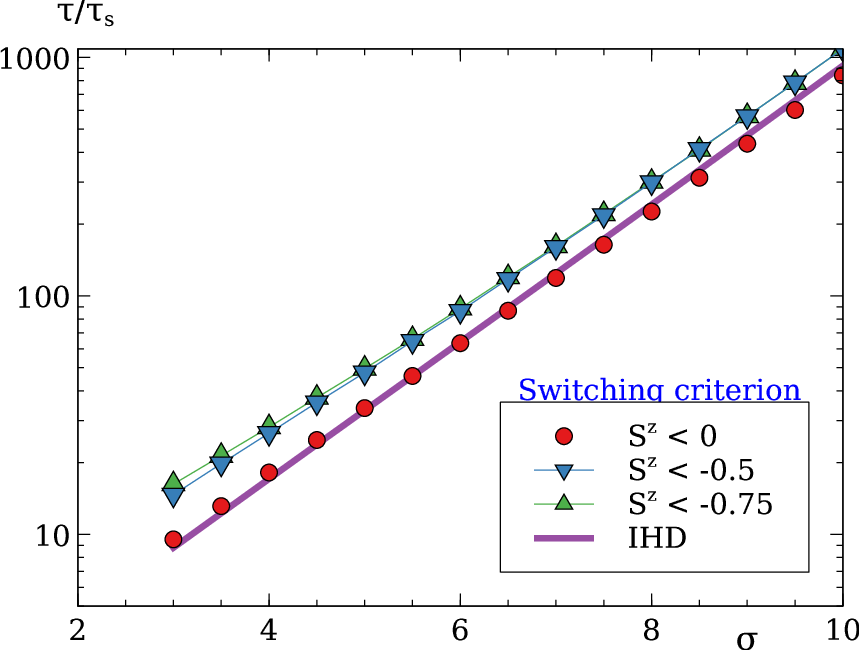}
\par\end{centering}
\caption{\label{fig:MC_SwitchingCriterion}Influence of the switching criterion on MC results
compared to the IHD formula results for noninteracting particles in a longitudinal
magnetic field: reduced relaxation time as a function of the reduced anisotropy $\sigma=KV/k_{B}T$.}
\end{figure}

In the present context, the relaxation time $\tau_{-}$ is the time for a 
magnetic moment to reverse from the $+z$ direction to the $-z$ direction. 
Therefore, in order to compute the relaxation time $\tau_{-}$,
one has to define the reversal criterion. While this is unambiguous
in analytical theory, in numerical simulations this requires some
care since the results may strongly depend on it and the potential
energy-scape studied. In our MC calculations we decided
that the reversal of $\bm{S}_{i}$ is achieved when its $z$ component
becomes, for the first time, smaller than some value $\bm{S}_{\mathrm{flip}}$.
For example, one may decide to choose $\bm{S}_{\mathrm{flip}}$
equal to $S_{z}$ at the saddle point, \emph{i.e.} $\bm{S}_{\mathrm{flip}}=S_{z}\left(\theta_{s}\right)=h$
(in the longitudinal case, there is a rotational symmetry of the uniaxial
potential and the saddle point is given by $\theta_{s}=\arccos(h)$,
for any azimuthal angle $\varphi$, so, there is an infinite number
thereof). However, this would underestimate the relaxation time because
the magnetic moment may return to its starting metastable state even 
if $S_{z}<S_{z}\left(\theta_{s}\right)$.
After several tests, we chose $S_{\mathrm{flip}}=0$. For each NM,
the number $n_{i}$ of MC steps for which its magnetic moment $\bm{S}_{i}$
reverses its direction is stored and the relaxation time $\tau_{-}$
is calculated according to
\[
\tau_{-}=\frac{1}{N}\sum_{i}n_{i}\delta t.
\]
In the absence of DI, our choice $S_{\mathrm{flip}}=0$ is
confirmed by the TQMC results shown in Fig. \ref{fig:MC_SwitchingCriterion} 
with different criteria, compared to the Aharoni-N\'eel-Brown (IHD) expression, see Eq.~(\ref{eq:NB-RR}) below.

To minimize statistical errors, the relaxation time $\tau_{-}$ was averaged over
100 simulations using parallel computing, with each simulation employing
a different random number sequence.

\section{Results and discussion\label{sec:results}}

We have studied chains of spherical NM of diameter $D=5\,\mathrm{nm}$
of iron with anisotropy constant $K\simeq2\times10^{5}\,\mathrm{J.m}^{-3}$
and saturation magnetization $M_{S}\simeq1.4\times10^{6}\,\mathrm{A.m}^{-1}$.
The damping parameter is $\alpha=1$.

\subsection{Relaxation rate}

\subsubsection{Langer's approach}

In order to compute the energy barrier with $\bm{h}=\pm h\bm{e}_{z}$,
we need to determine the saddle point. For this, we compute the functional
derivative $\delta\mathcal{E}_{i}/\delta S_{i,\alpha},\ \alpha=x,y,z$.
The two transverse components (\emph{i.e.} $\alpha=x,y$) yield the
constraints: $\sum_{j}V_{ij}S_{j}^{\alpha=x,y}=0$. This is consistent
with the fact that the problem is symmetric with respect to the $z$-axis.
The longitudinal component $S_{z}$ contains the most relevant information
about the energy barrier:
\begin{equation}
S_{i,z}=-h-4\xi\sum_{j}V_{ij}S_{j,z}\label{eq:phi_z}
\end{equation}
which can be self-consistently solved, leading to the following result
(to first order in $\xi$)
\begin{equation}
S_{i,z}^{\left(s\right)}=\cos\theta_{i}^{\left(s\right)}=-h\left[1-4\xi I_{i}\right],\label{eq:eta_saddle_longitu}
\end{equation}
with $I_{i}=\sum_{j}V_{ij}$ and where $\left(s\right)$ refers to
the saddle point.
We have checked that for chains of more than 20
particles, the sum $I_{i}$ is nearly constant along the chain with
a maximum deviation at the edges of less than 5\%. This implies that
the lattice sum $I$ can be considered as independent of the site
at which it is computed. Its limit is given by the Riemann zeta function
$\zeta\left(3\right)\approx1.202$. Henceforth, in a first approximation,
we consider chains that are sufficiently long to neglect edge effects,
such that $I_{i}=I$ and consequently $\theta_{i}^{\left(s\right)}=\theta_{s}$
(the limitations of this approximation for short chains, $N<20$, are discussed in a remark at the end of Sec.~\ref{sec:results}).
Substituting $S_{i,z}^{\left(s\right)}$ from Eq. (\ref{eq:eta_saddle_longitu})
back into Eq. (\ref{eq:EnergyLocal}) yields
\begin{equation}
\mathcal{E}_{s\parallel}^{\left(0\right)}\left(r\right)=\frac{h^{2}}{2}\left(1-4\tilde{\xi}\right).\label{eq:energy_saddle_longi}
\end{equation}
where we have introduced $\tilde{\xi}\equiv\xi I$ as the effective DI coefficient, with $I=\zeta(3)$ (the standard 1D
lattice sum factor), the upscript ${\left(0\right)}$ refers to the energy being computed precisely at the saddle point and not in its vicinity, as we shall see later.

The energy $\mathcal{E}_{-}^{\left(0\right)}$ at the (meta)stable state is obtained by inserting $S_{i,z}^{\left(\pm\right)}=\pm1$
into Eq. (\ref{eq:EnergyLocal}), upon which the energy barrier $\beta\Delta E_{\pm}$
with respect to the (meta)stable state becomes
\begin{equation}
\beta \Delta E_{\pm}^{\parallel}=\sigma\left[(1\pm h)^{2}+4\tilde{\xi}\left(1-h^{2}\right)\right]\label{eq:energy_barrier}
\end{equation}
where the $\pm$ sign refers to the relative orientation
of the field with respect to the magnetic moments. This approach does
not assume that all spins rotate in unison, but rather that a virtual
macro-spin has to jump over an energy barrier dressed by dipolar interactions
and uniaxial anisotropy.

Computing the partition functions in the vicinity of the saddle point
and metastable state is gained by performing a second-order expansion
of the energy. For this purpose, it is easier to rewrite the equation
of the energy in Eq. (\ref{eq:EnergyLocal}) in spherical coordinates
$(\theta,\varphi)$. Inserting the value of $\cos\theta_{s}$ obtained
in Eq. (\ref{eq:eta_saddle_longitu}) in the expression of the second
derivative and keeping only the linear terms in $\xi$, leads to
\begin{equation}
\mathcal{E}_{s}^{\parallel}\simeq\mathcal{E}_{s\parallel}^{\left(0\right)}+\frac{1}{2}\underbrace{\left(h^{2}-1\right)\left[1+\tilde{\xi}\frac{8h^{2}}{1-h^{2}}\right]}_{=-\lambda_{t}<0}\left(\theta-\theta_{s}\right){}^{2}.\label{eq:energy_at_saddle_point}
\end{equation}
Note that rotation invariance around the $z$-axis results in the
fact that $\mathcal{E}_{s}^{\parallel}$ does not depend on $\varphi$.
Furthermore, as stated above, we may neglect the edge effects for
chains longer than 20 particles. Hence, the partition function at
the saddle point can be factorized and easily evaluated, as a Gaussian
integral
\begin{equation}
\mathcal{Z}_{s,\parallel}=\left(2\pi\right)^{3/2}\sqrt{\frac{k_{B}T}{k}}\ e^{-\beta E_{s\parallel}^{\left(0\right)}}.\label{eq:partition_fct_saddle}
\end{equation}
To first order, the dipolar field is only present because it shifts
the energy of the saddle point by hardening the anisotropy. In the
absence of $\xi$ in Eq. (\ref{eq:partition_fct_saddle}), one recovers
the standard expression for a single spin with a uniaxial anisotropy.

Concerning $\mathcal{Z}_{m}$, the evaluation of the partition function
near the metastable state, the procedure remains the same. However,
since there is a rotational invariance around the $(Oz)$ axis, this
part is easier to compute by using the Cartesian coordinates. Therefore
the partition function is given by
\begin{equation}
\mathcal{Z}_{m} = e^{-\beta E_{-}^{\left(0\right)}}\frac{2\pi k_{B}T}{k\left(1-h\right)\left[1+\tilde{\xi}\frac{4}{1-h}\right]}.
\label{eq:partition_fct_metastable}
\end{equation}

From Eqs. (\ref{eq:partition_fct_saddle}) and (\ref{eq:partition_fct_metastable}),
we obtain the ratio $\mathcal{Z}_{s,\parallel}/\mathcal{Z}_{-}$
\begin{equation}
\frac{\mathcal{Z}_{s,\parallel}}{\mathcal{Z}_{m}}=\sqrt{\frac{2\pi k}{k_{B}T}}\ e^{-\beta\Delta E_{-}^{\parallel}}\ \left(1-h\right)\left[1+\tilde{\xi}\frac{4}{1-h}\right]\label{eq:ratio_ZsZ1}
\end{equation}
where the energy barrier $\Delta E_{-}^{\parallel}$ is given by Eq.
(\ref{eq:energy_barrier}).

The attempt frequency $\kappa$ is still missing to compute the relaxation
rate as given by the expression in Eq.~(\ref{eq:LangerRR}). $\kappa$
is given by the first nonzero negative eigenvalue of the transfer
matrix obtained after linearizing the Landau-Lifshitz equation. To
this end, we expand the coordinates $\left(\theta,\varphi\right)$
around the saddle point $\left(\theta_{s},\varphi_{s}\right)$, \emph{i.e.}
$\theta\simeq\theta_{s}+t,\varphi\simeq\varphi_{s}+p$ and next expand
the energy to second order (call the result $\mathcal{E}_{s}^{\left(2\right)}$)
in $t,p$ upon which the Landau-Lifshitz equation becomes
\[
\left\{ \begin{array}{lcl}
\overset{\cdot}{t} & = & -\partial_{p}\mathcal{E}_{s}^{\left(2\right)}-\alpha\partial_{t}\mathcal{E}_{s}^{\left(2\right)},\\
\\
\dot{p} & = & -\alpha\partial_{p}\mathcal{E}_{s}^{\left(2\right)}+\partial_{t}\mathcal{E}_{s}^{\left(2\right)}.
\end{array}\right.
\]
Where $\alpha$ is the damping parameter. The two equations above
can be recast into the following matrix form (using the notation $\eta_{i}=\left(t,p\right)$)
\[
\partial_{t}\eta_{i}=\sum_{j}M_{ij}\partial_{\eta_{j}}\mathcal{E}_{s}^{\left(2\right)}.
\]
Close to the saddle point, the energy may be expressed as $\mathcal{E}_{s}=\mathcal{E}_{s}^{\left(0\right)}+\frac{1}{2}\lambda_{t}t^{2}+\frac{1}{2}\lambda_{p}p^{2}$.
In the present case, we have $\lambda_{p}=0$ and $\lambda{}_{t}$
is defined in Eq. (\ref{eq:energy_at_saddle_point}). Hence, the eigenvalue
of the resulting matrix leads to
\begin{equation}
\left|\kappa\right|=\alpha\left(1-h^{2}\right)\left[1+\tilde{\xi}\frac{8h^{2}}{1-h^{2}}\right].\label{eq:attempt_longi}
\end{equation}
The result in Eq. (\ref{eq:attempt_longi}) and the ratio in Eq. (\ref{eq:ratio_ZsZ1})
are used in Eq. (\ref{eq:LangerRR}) to compute the relaxation rate
in longitudinal field $\Gamma_{-\to s}$. By simply performing the
change $h\to-h$, one can deduce the rate $\Gamma_{+\to s}$. Adding
up these two equations renders the total relaxation rate $\Gamma_{\parallel}=\Gamma_{-\to s}+\Gamma_{+\to s}$
of the chain's magnetic moment
\begin{align}
\Gamma_{\parallel}\left(h,\sigma,\xi\right) & =\frac{\sqrt{\sigma}}{\sqrt{\pi}\tau_{s}}\left(1-h^{2}\right)\left(1+8\tilde{\xi}\frac{h^{2}}{1-h^{2}}\right)\,\nonumber\\
&\times\left[\left(1-h+4\tilde{\xi}\right)e^{-\Delta\mathcal{E}_{-}^{\parallel}}+\left(1+h-4\tilde{\xi}\right)e^{-\Delta\mathcal{E}_{+}^{\parallel}}\right].\label{eq:overall_gamma_long}
\end{align}
where $\sigma$ is defined in Eq. \eqref{eq:sigma}, the reduced energy barriers $\Delta\mathcal{E}_{\pm}^{\parallel}$ in Eq. \eqref{eq:energy_barrier}, and $\tau_{s}=\frac{1+\alpha^{2}}{\alpha\gamma H_{a}}$.

It can readily be checked that in the absence of DI ($\xi=0$), this
expression recovers the well known Aharoni-Néel-Brown result \cite{brown63pr,Aharoni_PhysRev1969}
for longitudinal field, namely
\begin{align}
\Gamma_{\parallel}\left(h,\sigma,\xi=0\right) & =\frac{\sqrt{\sigma}}{\sqrt{\pi}\tau_{s}}\left(1-h^{2}\right)\times\nonumber\\
 & \left[(1-h)e^{-\sigma(1-h)^{2}}+(1+h)e^{-\sigma(1+h)^{2}}\right].\label{eq:NB-RR}
\end{align}

Eq. (\ref{eq:overall_gamma_long}) shows that the energy at the saddle
point changes due to the DI as well as to the external
magnetic field. The concomitant presence of the two contributions
leads to the additional cross term $\propto h^{2}\tilde{\xi}$, which
is symmetric with respect to the inversion of the applied magnetic
field. In the absence of the applied field, we have
\begin{equation}
 \Gamma_{\parallel}\left(0,\sigma,\xi\right) =\frac{2\sqrt{\sigma}}{\sqrt{\pi}\tau_{s}}e^{-\sigma\left(1+4\tilde{\xi}\right)}.
 \label{eq:GammaParal_h0}
\end{equation}

This shows that the DI contribution to the energy barrier depends
on the sign of $\tilde{\xi}$, or more precisely on that of the lattice
sum $I$, and thereby on the shape of the assembly. So, for a prolate
assembly $I>0$, the DI enhance the energy barrier, and vice versa.
The factor $1+4\tilde{\xi}$ may be absorbed in $\sigma$ or in the
anisotropy constant $k$ which then leads to an effective constant
$k^{\prime}=k\left(1+4\tilde{\xi}\right).$ This means that the chain
of interacting MN would behave as a macrospin with effective uniaxial
anisotropy of constant $k^{\prime}$ with an easy axis along the chain.
On the other hand, the result in Eq. (\ref{eq:GammaParal_h0}) shows
that the contribution of DI to the prefactor only occurs in the presence
of the magnetic field. This leads to a subtle competition between
the applied field and DI regarding the effect of damping on intra-well
relaxation.


\subsubsection{Comparison with MC results}

When DI are present, it is crucial to determine the minimum chain
length needed to accurately capture their effects. After running MC simulations
with varying $N$, we found that a chain of $10^{3}$
NM is sufficient to correctly model DI effects and obtain
reliable statistics. The results are reproducible across different
simulations with varying random number sequences.

\begin{figure}[H]
\begin{centering}
\includegraphics[scale=0.4]{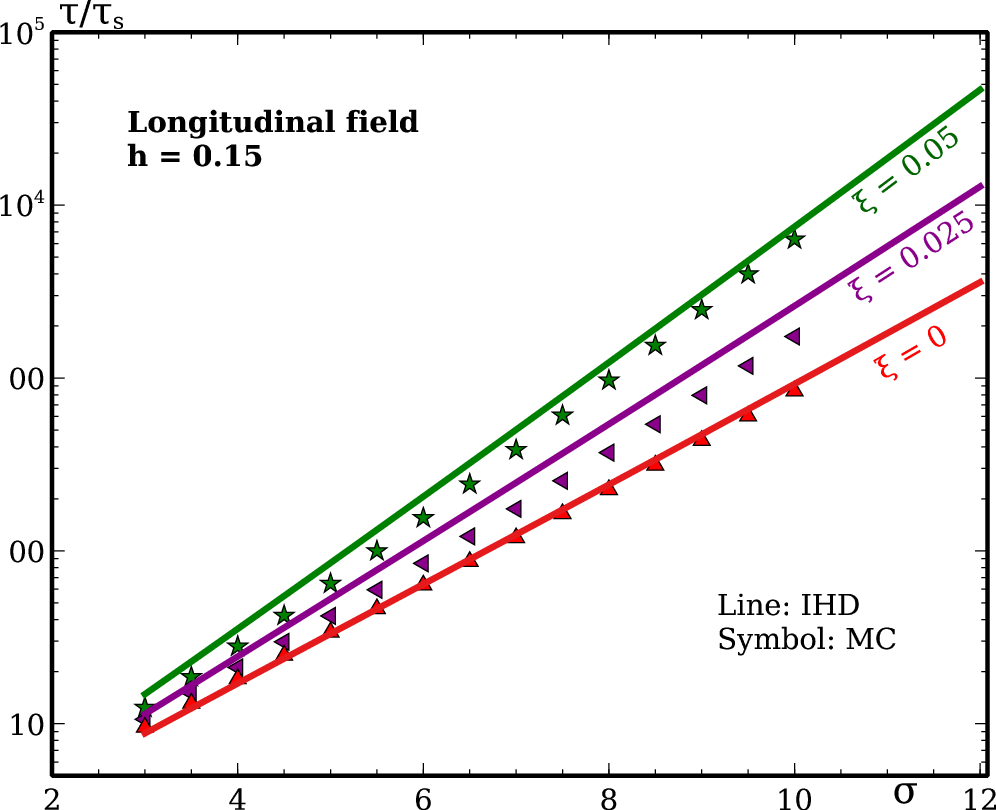}
\par\end{centering}
\caption{\label{fig:TauvsSig-IHDvsMC_longH015}Reduced relaxation time $\tau_{-}/\tau_{s}$
from Eq. (\ref{eq:overall_gamma_long}) as a function of $\sigma=KV/k_{\mathrm{B}}T$
for noninteracting ($\xi=0$) and interacting ($\xi\neq0$)
nanomagnets, in a longitudinal magnetic field ($h=0.15$). The solid
lines represent the IHD formula and symbols correspond to the MC simulations.}
\end{figure}

In Fig. \ref{fig:TauvsSig-IHDvsMC_longH015} we plot the (reduced)
relaxation time $\tau_{-}/\tau_{s}$, which is the inverse of $\Gamma_{-}$ in Eq. (\ref{eq:overall_gamma_long}), as a function of $\sigma=KV/k_{\mathrm{B}}T$
rendered by the IHD formula (continued lines) and MC simulations (symbols)
for $\xi=0,0.025,0.05$. As discussed after Eq. (\ref{eq:GammaParal_h0}) and
in connection with the result from Eq. (\ref{eq:energy_barrier})
for the effective energy barrier, the IHD approach predicts an increase
of the barrier in the presence of DI, and this is confirmed by the
MC calculations.

For a macrospin, the agreement between the IHD analytical approach,
the Langevin dynamics and the TQMC has already been demonstrated in
previous works \cite{ChubykaloEtAl_prb03}, in the IHD 
regime. Here, we show that this agreement applies, to some
extent, to the case of a chain of interacting magnetic moments.

\subsection{Chain magnetization dynamics}

As has been discussed by several authors \cite{cofetal95prb, gar96pre, garpal00acp}, the magnetization relaxation operates through several modes
each with its corresponding eigenvalue or relaxation rate. However,
it was shown in Ref. \cite{cofetal95prb} that the amplitude of higher-frequency
modes is very small and that the relaxation curve of the magnetization
can be modeled using a two-exponential function, corresponding to
the relaxation in the well and to overbarrier. More precisely, the magnetization's time dependence
$m(t) = \sum_{i=1}^{\mathcal{N}} S_i^{z}/\mathcal{N}$, can be written as
\begin{equation}
m\left(t\right)=A+Be^{-\Gamma_{\mathrm{b}}t}+Ce^{-\Gamma_{\mathrm{w}}t},\label{eq:m(t)_general}
\end{equation}
where $A,B,C$ are functions of temperature and effective field, and
$\Gamma_{\mathrm{b}}$ and $\Gamma_{\mathrm{w}}$ are the relaxation
rates related to over-barrier and intra-well fluctuations, respectively
($\varGamma_{w}\gg\varGamma_{b})$, with $\varGamma_{b}=\Gamma_{\parallel}$
being given by Eq. (\ref{eq:overall_gamma_long}). Note that, although
the number of relaxation modes grows with the chain length $\mathcal{N}$,
the perturbative smallness of $\xi$ ensures that the spectral gap between
$\Gamma_{\mathrm{w}}$ and $\Gamma_{\mathrm{b}}$ remains large, so
intermediate modes carry negligible weight and the two-exponential
truncation remains adequate. The coefficients $A,B$
and $C$ have to be determined with the help of several conditions, in the case of a longitudinal applied field:
\begin{itemize}
\item At initial time, \emph{i.e.} $t=0$, $m\left(0\right)=m_{0}$, which
is the magnetization of the initial state.
\item At asymptotically large times, $m\left(t\rightarrow\infty\right)$
must yield the equilibrium magnetization $m_{\mathrm{eq}}$.
\item In strong enough fields and/or low temperatures, or equivalently when
the effective energy barrier is high enough, the magnetic moments
stay most of the time in one minimum. This implies that the over-barrier
relaxation is suppressed\emph{, }i.e. $\Gamma_{\mathrm{b}} t\rightarrow 0$,
and the magnetization may be denoted by $m_{\mathrm{w}}$.
\end{itemize}
This leads to the following expression for the magnetization
\begin{equation}
m\left(t\right)=m_{\mathrm{eq}}+\left(m_{\mathrm{w}}-m_{\mathrm{eq}}\right)e^{-\Gamma_{\mathrm{b}}t}+\left(m_{0}-m_{w}\right)e^{-\Gamma_{\mathrm{w}}t}.\label{eq:m(t)_general-v2}
\end{equation}

Now, we need explicit expressions of $m_{0},m_{\mathrm{eq}}$ and
$m_{\mathrm{w}}$.
$m_{0}$ is fixed at the beginning and so it is determined by the
initial state. In the present calculations, the magnetic field is
set at $-\bm{e}_{z}$ and the initial magnetic moments at $\bm{e}_{z}$.
Hence, $m_{0}=1$, while its projection onto
the magnetic field is $-1$. The other magnetizations depend on the
situation.

\subsubsection{Noninteracting chains}

The equilibrium magnetization for a noninteracting magnetic moment in a longitudinal field 
is given by\cite{garpal00acp}
\begin{equation}
m_{\mathrm{eq}}^{\left(0\right)}\left(h,\sigma\right)=\frac{e^{\sigma}}{\sigma}\frac{\sinh(2\sigma h)}{Z_{\parallel}}-h,\label{mag_parallel}
\end{equation}
where $Z_{\parallel}$ is the partition function 
\begin{equation}
Z_{\parallel}=\int_{-1}^{1}dz\,e^{\sigma z^{2}+2\sigma hz}.\label{Z_parallel}
\end{equation}

Similarly, $m_{\mathrm{w}}$ is defined by an integration in Eq. (\ref{Z_parallel})
from \textit{h} to 1, \emph{i.e.} corresponding to the upper well
(the DC field is along $-{\bf e}_{z}$), and is computed numerically.

Next,

\begin{equation}
 \Gamma_{\mathrm{W}}=\frac{1}{\tau_{s}}\left(1-h\right)\label{eq:Gamma_well}
\end{equation}
is the relaxation rate in the metastable state and $\Gamma_{\mathrm{b}}=\Gamma_{\parallel}\left(h,\sigma,0\right)$,
see Eq. (\ref{eq:NB-RR}).

\subsubsection{Interacting chains}

Next, in the presence of DI,
\begin{equation}
 \Gamma_{\mathrm{W}}=\frac{1}{\tau_{s}}\left(1-h+2\tilde{\xi}\right)\label{zq:Gamma_well_DI}
\end{equation}
is the relaxation rate in the metastable state.

In Ref. \cite{sabsabietal13prb} we derived the following asymptotes
\begin{equation}
m_{\mathrm{eq}}^{\mathrm{LF}}\simeq\left\{ \begin{array}{lll}
m^{\left(0\right)}\left[1+\tilde{\xi}\left(\frac{1}{3}+\frac{4}{45}\sigma\right)\right], &  & \sigma\ll1,\\
\\
m^{\left(0\right)}\left[1+\tilde{\xi}\left(1-\frac{1}{\sigma}\right)\right], &  & \sigma\gg1,
\end{array}\right.\label{eq:mag_lowfield}
\end{equation}
for low fields.

To preserve the double-well nature of the energy potential, the field
is small while the effective energy barrier $\sigma$ is usually $\gg1$.
Therefore, the equilibrium magnetization in Eq. (\ref{eq:m(t)_general-v2})
is given by
\begin{equation}
m_{\mathrm{eq}}\left(h,\sigma,\xi\right)\simeq m_{\mathrm{eq}}^{\left(0\right)}\left(h,\sigma\right)\left[1+\tilde{\xi}\left(1-\frac{1}{\sigma}\right)\right]\label{eq:meq_lowfield}
\end{equation}
and similarly for $m_{\mathrm{w}}\left(h,\sigma,\xi\right)$.

\begin{figure}[H]
\begin{centering}
\includegraphics[scale=0.35]{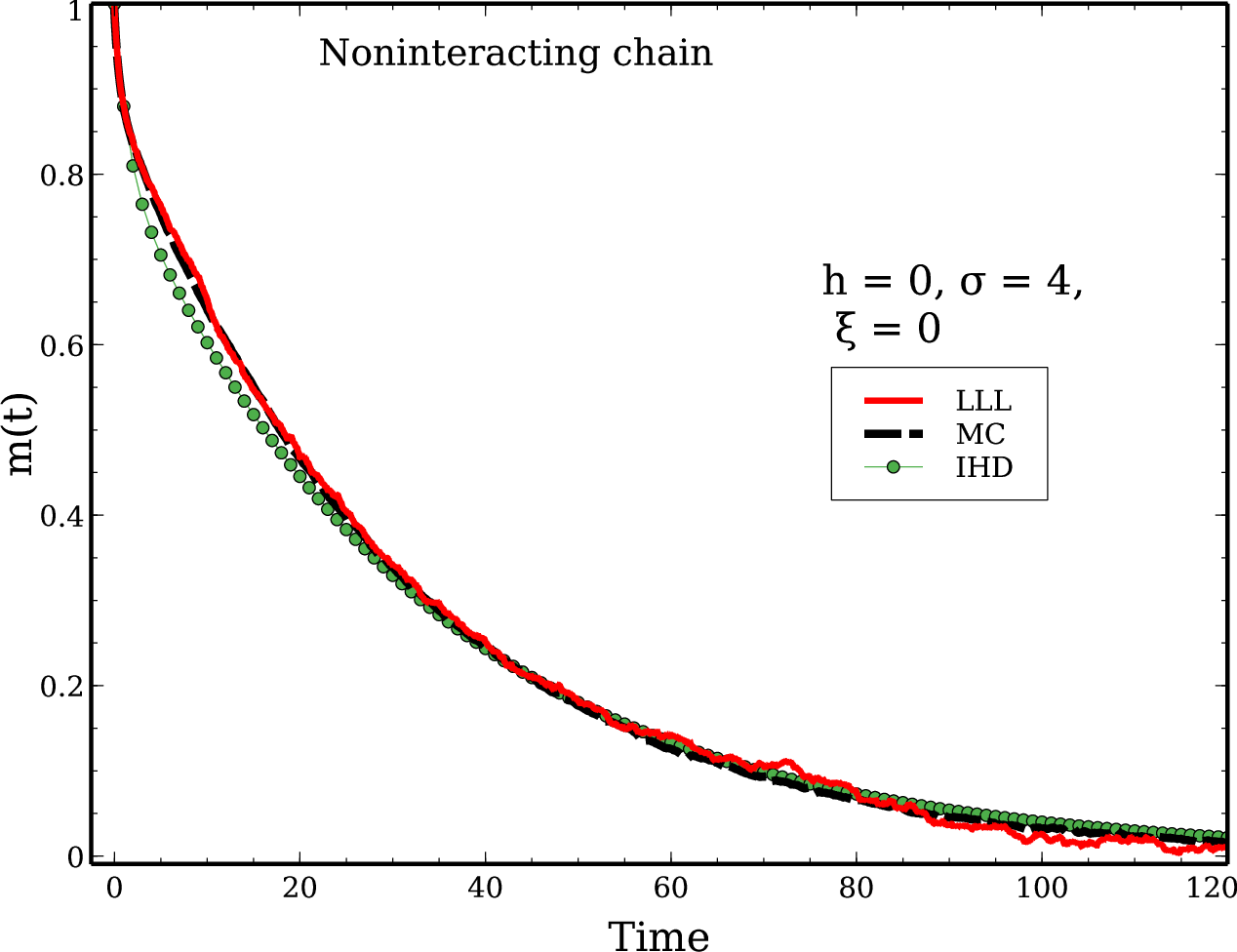}
\par\end{centering}
\caption{\label{fig:mt-IHD-LLL-MC}Magnetization relaxation rendered by MC,
LLL and analytical expression (\ref{eq:m(t)_general-v2}) for noninteracting
particles ($\xi=0$), in zero field ($h=0$), for a reduced anisotropy $\sigma=4$. The MC curve was obtained by averaging over 50 realizations to reduce statistical errors.}
\end{figure}

In Fig. \ref{fig:mt-IHD-LLL-MC}, we compare the time-dependent magnetization obtained by 
three methods: IHD approximation, MC, and Landau-Lifshitz-Langevin dynamics (LLL),
in the noninteracting case and zero magnetic field. The results show good agreement over the whole relaxation time interval, and thus
the IHD approach provides a good approximation even in the unfavorable
regime of small $\sigma$ (here $4$). 
Having an analytical expression for this magnetization dynamics represents a distinct advantage for an 
easy first interpretation or fitting of experiments in the field. Using the formula obtained by the present IHD approach
is now straightforward, in contrast to TQMC, which requires coding 
skills and access to significant CPU time to treat NM interacting assemblies.

\paragraph{Effect of $\sigma$}
\begin{figure*}[h!]
\begin{centering}
\includegraphics[scale=0.45]{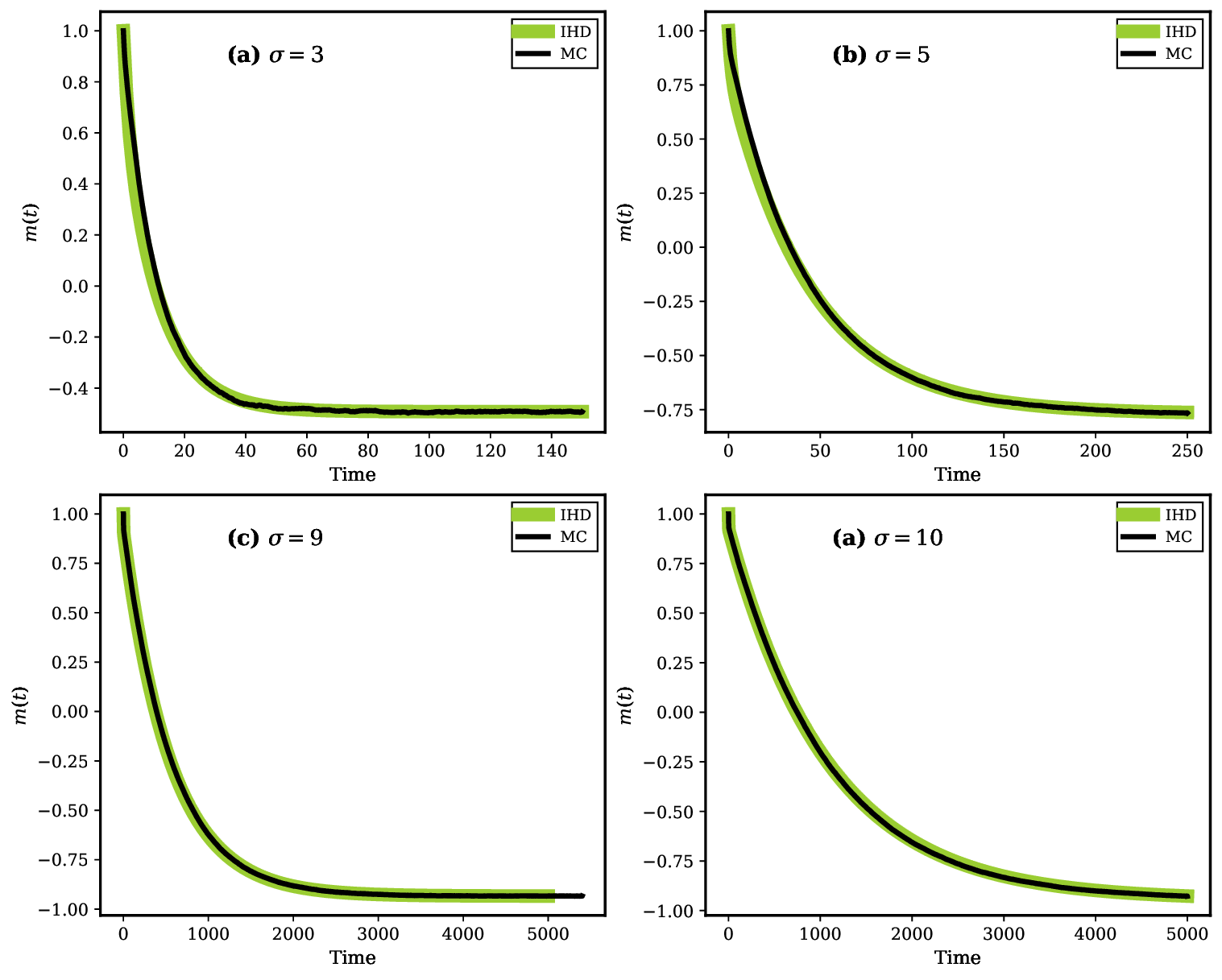}
\par\end{centering}
\caption{\label{fig:mt-IHD-MC_h015xi0_varsig}Magnetization relaxation rendered
by MC and the IHD-based expression (\ref{eq:m(t)_general-v2}) for noninteracting
particles ($\xi=0$), with a fixed reduced field $h=0.15$ and varying the reduced anisotropy $\sigma=KV/k_{B}T$.}
\end{figure*}

In Fig. \ref{fig:mt-IHD-MC_h015xi0_varsig}, we compare the MC and
IHD results for various values of $\sigma$, for the noninteracting
chain, in a magnetic field of amplitude $h=0.15$. The two
methods show good agreement for the whole time interval and all values
of $\sigma$, which improves even further for larger values of the latter.
\begin{figure*}[h!]
\begin{centering}
\includegraphics[scale=0.48]{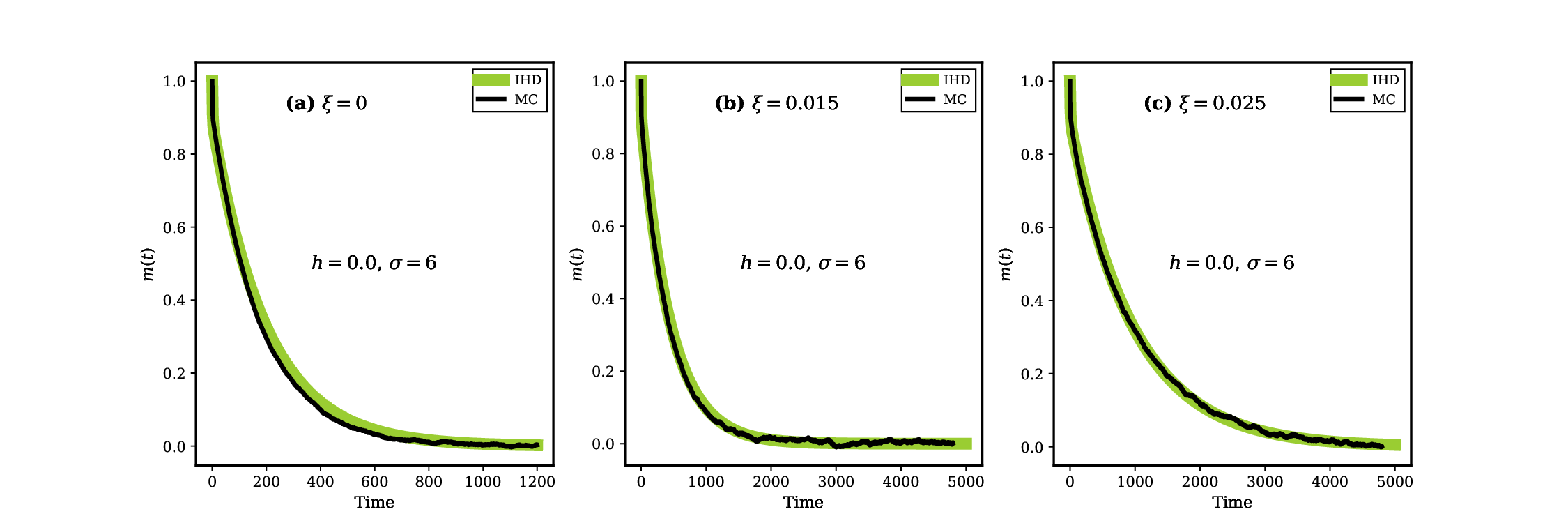}
\par\end{centering}
\begin{centering}
\includegraphics[scale=0.48]{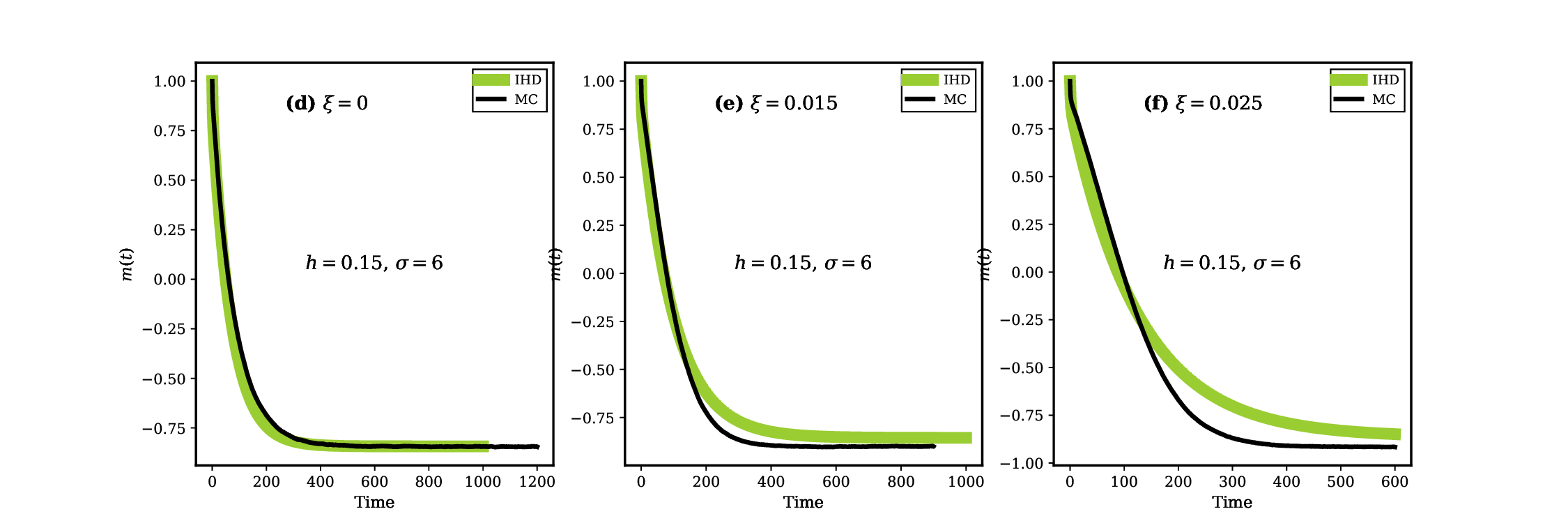}
\par\end{centering}
\caption{\label{fig:mt-IHD-MC_h0-h015-sig6_varxi}Magnetization relaxation
rendered by MC and the IHD-based expression (\ref{eq:m(t)_general-v2})
for interacting chains ($\xi\ne 0$), with a reduced anisotropy $\sigma=6$ and varying the DI parameter $\xi$. Top row (a--c): $h=0.0$; bottom row (d--f): $h=0.15$.}
\end{figure*}

\paragraph{Effect of $h$ and $\xi$}

Next, in Fig. \ref{fig:mt-IHD-MC_h0-h015-sig6_varxi}, we plot $m(t)$
as rendered by the IHD-based (semi-)analytical expression (\ref{eq:m(t)_general-v2})
and MC, for varying $\xi$, and two values of the applied field $h$,
and $\sigma=6$.

We see that, in the presence of either the magnetic field or the dipolar
field alone, the agreement between the IHD and MC results is quite
good, as seen in Fig. \ref{fig:mt-IHD-MC_h0-h015-sig6_varxi}(b-d). However, when both fields are simultaneously present, a discrepancy
emerges as $\xi$ increases (see the plots in Fig. \ref{fig:mt-IHD-MC_h0-h015-sig6_varxi}(f) for $h = 0.15$ and $\xi$ =
0.025). This arises because the IHD approach involves two successive
approximations related to the dipolar field. The first approximation
is made in the derivation of the relaxation rate expression, while
the second appears in the equilibrium magnetization of the interacting
chain, given by Eq. (\ref{eq:meq_lowfield}), which is obtained through
a perturbative expansion with respect to the dipolar field parameter
$\xi$. Nevertheless, the IHD formulation still provides an overall
good approximation of the MC results, as illustrated in Figs. \ref{fig:mt-IHD-MC_h0-h015-sig6_varxi}
(e - f).

{The present IHD approach focuses on weak DI, at first order in $\xi$, which successfully
describes the dynamics up to a value that depends on the applied field $h$ ($\xi\simeq 0.025$ for $h =0$ and $\xi\simeq 0.01$ for $h = 0.15$), 
in good agreement with TQMC (see, e.g., Figs.~\ref{fig:TauvsSig-IHDvsMC_longH015}
and \ref{fig:mt-IHD-MC_h0-h015-sig6_varxi}).  Extending the treatment to stronger DI would require a considerably
more involved calculation, as the saddle-point energy landscape itself is modified; in that regime a numerical
approach such as TQMC, as used here, is the natural recourse. The value of the present analytical approach lies
precisely in its simplicity: it provides practical, inexpensive closed-form expressions that already capture the
essential relaxation mechanism. Pushing the expansion to higher order -- e.g., via fluctuations and correlation
functions in the spirit of a Ginzburg-type criterion -- remains a possible route for the interested reader, but
at the cost of the simplicity that makes the present level of approximation useful.

\paragraph{Further discussion of the comparison between IHD and MC}

The results in Figs. \ref{fig:mt-IHD-MC_h015xi0_varsig} and \ref{fig:mt-IHD-MC_h0-h015-sig6_varxi}
show that the magnetization relaxation curves, and in particular the
short-time fast relaxation, rendered by MC, are well captured by the
(semi-)analytical expression (\ref{eq:m(t)_general-v2}) with the
two-exponential behavior. This indicates that the individual magnetic
moments first relax in their starting potential wells and next, owing
to DI, the collective mode relaxes more slowly over the DI-modified
energy barrier with the relaxation rate $\Gamma_{\parallel}\left(h,\sigma,\xi\right)$.
However, the IHD approach assumes that the switching occurs via
a saddle point that is a quasi-uniform state; see Eq. (\ref{eq:eta_saddle_longitu}) and discussion \textit{et seq.}

The fact that the discrepancy increases when both $h$ and $\xi$
are finite may also signal nonuniform switching channels beyond the
virtual-macrospin picture used in Langer's approach. Indeed, for
a longitudinally magnetized chain, each moment experiences the applied
field $h$ and a dipolar field. In a uniformly magnetized
state, all $\bm{S}_{i}$ are parallel, so the sign of each term in
the dipolar field depends only on the direction of ${\bf e}_{ij}$,
see Eq. (\ref{eq:DItensor}). As such, sites at the center see nearly
symmetric neighbors on both sides; the positive and negative contributions
partly cancel, yielding a smaller net destabilizing transverse component
but a stronger stabilizing (longitudinal) one; sites at the \emph{edges}
lack one side of neighbors and thus a smaller stabilizing dipolar
field. Indeed, a convenient way to summarize the spatially varying
dipolar stabilization is via the local (dimensionless) bias\cite{ManishEtAl_prb19}
\begin{equation}
h_{{\rm eff}}(i)\;\equiv\;h_{\mathrm{AZ}}\;+\;\xi\,\Lambda_{i}(N),\qquad \Lambda_{i}(N)=\sum_{n=1}^{i-1}\frac{1}{n^{3}}+\sum_{n=1}^{N-i}\frac{1}{n^{3}},\label{eq:heff}
\end{equation}
where $h_{\mathrm{AZ}}$ comprises the magnetic field and anisotropy field. Hence, we see that for a uniformly magnetized
chain (all spins parallel), $\Lambda_{i}(N)$ is largest near the chain
center and smallest at the edges. In the $N\to\infty$ limit,
\begin{equation}
\Lambda_{{\rm center}}\approx2\,\zeta(3),\qquad \Lambda_{{\rm edge}}\approx\zeta(3),\label{eq:S_Ninf}
\end{equation}
so the center experiences roughly \emph{twice} the stabilizing dipolar
bias that an edge site does.
For instance, with $\xi=0.025$ and $h=0.15$, the local energy barriers
computed from Eq.~(\ref{eq:energy_barrier}) with $\tilde{\xi}_{\rm edge}=\xi\,\zeta(3)$
and $\tilde{\xi}_{\rm center}=2\xi\,\zeta(3)$ yield a ratio
$\Delta\mathcal{E}_{\rm edge}/\Delta\mathcal{E}_{\rm center}\approx 0.93$,
confirming that the edge barrier is appreciably softer even for weak DI.

Consequently, the local escape barrier is larger (``stiffer'') in
the center and smaller (``softer'') near the edges. Then, when a
finite longitudinal field $h$ breaks the symmetry and $\xi$ increases,
the contrast in local stabilization grows, and reversal proceeds by
\emph{nucleation at the edges} followed by a front (domain-wall--like)
propagation, i.e.\ a nonuniform mode. Therefore, although DI 
are intrinsically long-ranged and each magnetic moment
interacts with all others, the finite geometry of the chain leads
to an important \emph{asymmetry} between sites since the sum $\Lambda_{i}$
depends on the site index $i$ even though every spin interacts with
all others; the difference originates from geometric asymmetry, not
from a cutoff of the dipolar range.
In the infinite-chain limit $N\!\to\!\infty$, the fraction of sites affected by edge effects vanishes, and almost all sites see
$\Lambda_{i}\to2\zeta(3)$; however, the two boundary sites retain $\Lambda_{i}\simeq \zeta(3)$
for any finite $N$.

To sum up, for $h=0$, the two escape channels are equivalent; a uniform
(macrospin-like) relaxation is competitive. For $h\neq0$, the metastable
well is biased, and the system prefers to start the reversal where
the \emph{local} barrier is smallest. From \eqref{eq:heff}--\eqref{eq:S_Ninf},
increasing $\xi$ amplifies the spatial gradient of the stabilizing
dipolar bias, raising the barrier less at the edges than at the center, thereby widening the barrier contrast.
Once the edge barrier falls below the (effective)
cost of forming and launching a reversal front, edge nucleation wins.
In practice, this crossover is facilitated because (i) increasing $\xi$
raises the energy barrier less at the edges than in the bulk, producing
a barrier contrast $\Delta\mathcal{E}_{\rm center}-\Delta\mathcal{E}_{\rm edge}
\propto\sigma\xi\,\zeta(3)(1-h^{2})$ that grows with both $\sigma$ and $\xi$, and (ii) the applied field
$h>0$ cooperates with this dipolar inhomogeneity to destabilize the edge first.

In MC results, this appears as a faster initial drop of $m(t)$ than
predicted by a single (uniform) over-barrier mode, followed by a slower
tail---a hallmark of two-stage (nucleation--propagation) dynamics.
In IHD, the same physics is mimicked by the two-exponential structure
with distinct rates and a (slightly) edge-biased effective prefactor
when $\xi$ grows.

To further elucidate the spatially non-uniform reversal mechanism suggested by the discrepancy between IHD and MC results when both \(h\) and \(\xi\) are finite, we directly analyze the temporal evolution of the correlation length \(\lambda(t)\) extracted from the equal-time spin-spin correlation function obtained by MC simulations:
\begin{equation}
G(r,t) = \frac{1}{N-r} \sum_{i=1}^{N-r} \left[ S_i^z(t) - m(t) \right] \left[ S_{i+r}^z(t) - m(t) \right]
\label{eq:GF}
\end{equation}
with \(\mathcal{N} = 1000\) being the chain length and \(r\) the inter-spin distance.
%

%

The correlation length \(\lambda(t)\) is obtained by fitting \(G(r,t)\) to an exponential decay \(G(r,t) \sim e^{- r / \lambda(t)}\) for each time step.

\begin{figure}[htbp]
\centering
\includegraphics[width=0.95\linewidth]{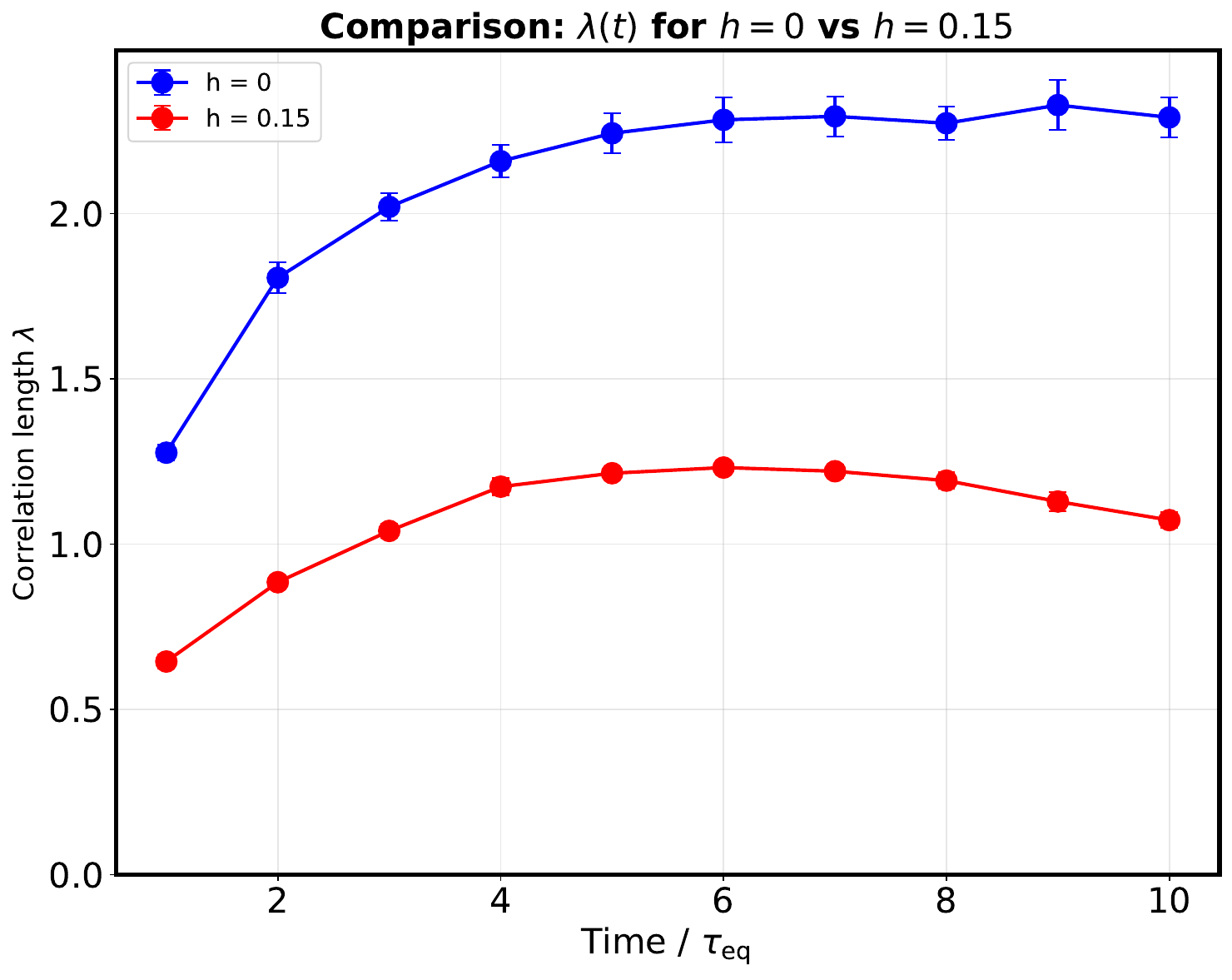}
\caption{Comparison of correlation length \(\lambda(t)\) for zero field \(h = 0\) (blue) and finite field \(h = 0.15\) (red) at a reduced anisotropy \(\sigma = 6\) and DI \(\xi = 0.025\). $\tau_\mathrm{eq}$ is the time at which equilibrium is reached. Error bars show fitting uncertainties. The zero-field case exhibits monotonic growth characteristic of uniform reversal, while the finite-field case shows a non-monotonic peak-and-decay pattern indicative of edge-nucleation dynamics.}
\label{fig:xi_comparison}
\end{figure}

Analysis of the results in Fig.~\ref{fig:xi_comparison} first shows that the correlation length is smaller in the presence of a (longitudinal) magnetic field. Then, comparing \(\lambda(t)\) for \(h = 0\) (blue) and \(h = 0.15\) (red), at \(\sigma = 6\) and \(\xi = 0.025\), reveals fundamentally different reversal mechanisms:
\begin{itemize}
    \item For \(h = 0\), \(\lambda(t)\) grows monotonically, which confirms our earlier conclusion that, in this time interval, the reversal operates via a quasi-uniform (macrospin-like) mode where all sites reverse cooperatively.
    \item For \(h = 0.15\), \(\lambda(t)\) displays a non-monotonic behavior: an initial increase to a maximum at intermediate times, followed by decay. This peak-and-decay pattern directly evidences edge-nucleation dynamics, as predicted by the spatial inhomogeneity of dipolar stabilization discussed in Eqs.~\eqref{eq:heff}--\eqref{eq:S_Ninf}. The initial rise corresponds to the nucleation and initial propagation of a reversal front from the edges, which increases spatial correlations. The subsequent decay reflects the completion of reversal and the suppression of correlations in the field-aligned state.
\end{itemize}

The distinct \(\lambda(t)\) behaviors confirm that the reversal mechanism crosses over from uniform to edge-nucleation propagation as the magnetic field is increased, explaining the observed relative discrepancies between IHD (which assumes quasi-uniform reversal) and MC simulations for finite \(h\) and \(\xi\).
Notably, the timescale at which \(\lambda(t)\) peaks in the \(h=0.15\) curve coincides with the temporal window where the IHD and MC magnetization curves in Fig.~\ref{fig:mt-IHD-MC_h0-h015-sig6_varxi}(f) diverge most, further connecting the onset of edge-nucleation dynamics to the observed discrepancy.

\smallskip
\noindent \textbf{Remark (Finite-size effects):} This is a remark about finite-size effects, e.g. when $N < 20$. The analytical model derived in Eq. \eqref{eq:energy_barrier} relies on the approximation of the lattice sum $I_i \simeq \zeta(3)$, which is valid for the bulk of long chains. In short chains, the spatial variation of the dipolar field becomes dominant. The spins at the chain ends ($i=1, N$) have only one neighbor, reducing the local stabilizing dipolar field ($I_{end} \simeq 1.2$) compared to the center ($I_{bulk} \simeq 2.4$). Consequently, the energy barrier is significantly lower at the edges, promoting edge nucleation even at low temperatures. While the global macrospin approximation fails for small $N$, our analytical framework can still predict the onset of reversal by substituting the global sum $I$ with the local value $I_{end}$ in the energy barrier expression. This correctly predicts that short chains are magnetically softer and reverse via an edge-mode instability, consistent with the ``edge-nucleation'' regime identified in our TQMC simulations.

Finally, we now discuss a related issue concerning the overall comparison
between the IHD and MC results and the time-scale consistency. As we have seen, the
two approaches describe the same relaxation dynamics but in slightly shifted
temporal frameworks, which are reconciled by introducing an empirical
scaling factor (between 0 and 1). This factor effectively accounts
for the dependence of the MC time step, or the radius $R$ of the
sphere used in the MC time step (\ref{eq:Deltat_MC}), on
the local effective field acting on each magnetic moment. When an
external field is applied along the chain axis, the dipolar field
competes with the combined anisotropy and Zeeman contributions, altering
the energy landscape and thus the characteristic relaxation time.
Consequently, the apparent time rescaling between IHD and MC simulations
reflects the field- and interaction-dependent modification of the
microscopic dynamics rather than a simple numerical artifact. In this
regard, it would be natural to extend the developments proposed in
Refs. \cite{nowaketal00prl,ChubykaloEtAl_prb03} to derive a generalized
expression for $R$ as a function of the effective field; on dimensional
grounds one expects $R^{2}\propto 1/[\sigma(1-h_{\mathrm{eff}})]$
(cf.\ Eq.~(\ref{eq:cone-radius})), where $h_{\mathrm{eff}}$ incorporates
both the Zeeman and dipolar contributions.

\section{Conclusion\label{sec:conclusion}}

We have developed a comprehensive analytical framework for magnetization relaxation in dipolar-coupled nanomagnetic chains, validated by TQMC simulations. The key achievements are:

1. \textbf{Analytical relaxation rate:} Equation~\eqref{eq:overall_gamma_long} provides a closed-form expression for the relaxation rate incorporating dipolar interactions via $\tilde{\xi}$, reducing to known results for noninteracting particles.

2. \textbf{Semi-analytical magnetization dynamics:} Equation~\eqref{eq:m(t)_general-v2} gives a two-exponential formula for the time dependence of the magnetization (\textit{i.e.}, $m(t)$) with explicit expressions for equilibrium and metastable magnetizations in interacting chains.

3. \textbf{Systematic validation:} Good agreement between analytical expressions and TQMC simulations across parameters $h$, $\sigma$, and $\xi$ confirms the perturbative treatment of dipolar interactions.

4. \textbf{Reversal mechanism crossover:} Correlation length analysis reveals a field-controlled crossover from quasi-uniform reversal ($h=0$) to edge-nucleation propagation ($h>0$), explaining discrepancies between analytical calculations and simulations at finite $h$ and $\xi$.

This study has rendered practical analytical expressions for the relaxation rate and magnetization relaxation curves which, to some extent, may help gain a clear picture of the assembly dynamics without resorting to extensive simulations.
The perturbative treatment is expected to break down when $4\tilde{\xi}\sim\mathcal{O}(1)$, \emph{i.e.}
$\xi\gtrsim 1/[4\zeta(3)]\approx 0.2$, at which point higher-order
DI corrections and multi-mode reversal channels can no longer be neglected.
Future work could extend to higher dimensions, complex anisotropies, and alternating fields for hyperthermia applications.

\section*{Acknowledgments}
D. Ledue acknowledges the Centre Régional Informatique et d'Applications Numériques de Normandie (CRIANN), where simulations were performed as Project No. 2022013. F. Vernay and H. Kachkachi acknowledge financial support from ANR-21-CE09-0043-01.


\end{document}